\documentclass[28pt]{article}

\usepackage[utf8]{inputenc}
\usepackage[T1]{fontenc}
\usepackage[english]{babel}
\usepackage{graphicx}
\usepackage{amsmath}
\usepackage{amsfonts}
\usepackage{booktabs}
\usepackage{nicefrac}
\usepackage{microtype}
\usepackage{adjustbox}
\usepackage{caption}
\usepackage{longtable}
\usepackage{lipsum}
\graphicspath{ {./images/} }

\usepackage[authoryear]{natbib}

\usepackage{hyperref}
\hypersetup{
    colorlinks=true,
    linkcolor=black, 
    citecolor=black,
    urlcolor=blue,
    pdfborder={0 0 0} 
}\newpage

\renewcommand{\and}{\&}

\title{Topic-Enriched Embeddings to Improve Retrieval Precision in RAG Systems for LLMs}

\author{
 Rodrigo Kataishi, Ph.D. \\
  CONICET Research Fellow\\
  National University of Tierra del Fuego\\
  Ushuaia, Argentina \\
  \texttt{rkataishi@untdf.edu.ar} \\
}
\date{}

\begin{document}
\maketitle

\begin{abstract}
Retrieval-Augmented Generation (RAG) systems rely on document retrieval to ground large language model outputs, yet retrieval quality often degrades in corpora where topics overlap and relevant evidence is distributed across long, heterogeneous texts. This paper proposes topic-enriched embeddings, a hybrid representation that integrates term-frequency signals (TF-IDF), dimensionality-reduced semantic structure (LSA), and probabilistic topic mixtures (LDA) into a unified vector space anchored by contextual sentence embeddings (\texttt{all-MiniLM-L6-v2}). The approach injects corpus-level thematic information into dense representations through two fusion strategies, concatenation and weighted averaging, while preserving computational tractability via latent-space compression. Empirical evaluation on a legal corpus of 12,436 documents related to Argentina’s Law 19.640 shows that topic enrichment improves both clustering coherence and retrieval effectiveness relative to statistical, probabilistic, and contextual-only baselines, with consistent gains in Precision@k, Recall@k, and F1. The results suggest that explicitly incorporating latent topic structure into embedding construction can reduce redundant or off-topic chunk retrieval, strengthening the evidential grounding of RAG pipelines in knowledge-intensive settings.

\end{abstract}

\vspace{30pt}
\newpage

\section{Introduction}
Retrieval-Augmented Generation (RAG) systems have enabled the ability of large language models (LLMs) to perform precise knowledge-intensive tasks by integrating external information through document ingestion \citep{lewis2021retrievalaugmented, zhao2024retrievalaugmented}. At the heart of these systems lies the retrieval mechanism, which determines the relevance of document chunks to user queries and uses that data as the context for LLM responses. Yet, retrieval precision continues to be a critical bottleneck, particularly when working with large datasets that exhibit heterogeneous topics and high thematic diversity \citep{beltagy2020longformer}. These issues often lead to the retrieval of irrelevant or redundant chunks, undermining the precision and reliability of downstream generation tasks \citep{finardi2024chronicles}.

Despite recent advances in context window expansion, retrieval remains one of the most critical and underexplored challenges in modern LLMs  \citep{google2024gemini}. Even with larger attention spans and extended context lengths, empirical studies have shown that retrieval systems often suffer from “lost in the middle” effects—where relevant information placed between less salient content is ignored or diluted—ultimately compromising model accuracy \citep{meta2025llama4}. This phenomenon increases the risk of hallucinations and misaligned answers in RAG applications, particularly when dealing with dense regulatory or technical corpora. As a result, enhancing retrieval fidelity has become a priority for improving factual grounding and minimizing semantic drift in LLM-based pipelines.

While high-performing open-source SOTA embedding models—such as \texttt{nomic-embed-text}, \texttt{mxbai-embed-large}, \texttt{bge-m3}, \texttt{snowflake-arctic-embed}, and especially \texttt{all-MiniLM-L6-v2}—have significantly advanced semantic retrieval, they tend to operate as stand-alone solutions. In most applications, they prioritize fine-grained semantic similarity, yet overlook thematic coherence and structural ordering across large, heterogeneous corpora. Nonetheless, recent work shows that traditional bag-of-words methods remain competitive across a wide range of classification challenges \citep{graff2025bow}, reinforcing our decision to retain TF-IDF in the embedding ensemble. This gap becomes particularly evident in domains that require precise retrievals—such as legal or regulatory texts—and where disambiguation of overlapping topics, hierarchical relationships among concepts, or deeper contextual grounding is essential. In this paper, we argue that a more robust solution lies not in replacing these models, but in reconfiguring them. The paper proposes a hybrid retrieval strategy that integrates the topic-level resolution capacity of statistical and probabilistic methods with the semantic depth of contextual embeddings, specifically leveraging the open-source \texttt{all-MiniLM} model. This layered architecture constitutes the methodological innovation of our approach and provides a functional enhancement to how RAG systems structure, rank, and retrieve complex textual content.

This research proposes topic-enriched embeddings, a refined approach that integrates traditional statistical models and probabilistic topic modeling with modern contextual embeddings \citep{deerwester1990indexing, blei2003latent}. This method captures both term-level and topic-level semantics, addressing the retrieval challenges posed by complex datasets \citep{li2022hierarchical}. By enriching embeddings with latent topic structures and leveraging dimensionality reduction, topic-enriched embeddings enable more precise document clustering and retrieval.

Dimensionality reduction plays a pivotal role in this approach. Techniques such as LSA preserve critical semantic features while minimizing computational overhead \citep{deerwester1990indexing}, ensuring that the method remains scalable for large datasets. This balance of precision and efficiency is particularly valuable for RAG systems, where effective retrieval underpins high-quality response generation \citep{zhao2024retrievalaugmented}.

The practical applicability of topic-enriched embeddings is demonstrated through empirical validation using a legal text dataset, and this approach significantly improves clustering coherence and retrieval precision \citep{huseynova2024enhanced}. Although the method introduces additional computational steps, the results justify the trade-off: accuracy in retrieval takes precedence over computational cost, as well-defined processing steps enhance the overall quality of the system. By reducing the retrieval of irrelevant or redundant chunks, it addresses core inefficiencies in existing systems, ensuring more accurate grounding for downstream generation tasks \citep{lu2022precise}. These results underscore the utility of topic-enriched embeddings in advancing knowledge-intensive RAG systems \citep{finardi2024chronicles}.

Additionally, the method emphasizes reproducibility and scalability. An open-source framework for embedding enrichment is provided, ensuring adaptability to diverse corpora and retrieval-intensive applications. This flexibility reinforces the relevance of the proposed approach across a wide range of domains, paving the way for future advancements in RAG technologies.

The manuscript is organized as follows. Section~\ref{sec:related} reviews prior work on topic modeling, embedding techniques, and retrieval-augmented generation (RAG). Section~\ref{sec:method} introduces the proposed embedding enrichment method. Section~\ref{sec:experiments} describes the dataset, preprocessing steps, baseline methods, and evaluation framework. Section~\ref{sec:results} presents the experimental findings, while Section~\ref{sec:discussion} examines their implications and limitations. Finally, Section~\ref{sec:conclusion} summarizes the main contributions and outlines directions for future research.

\section{Related Work}
\label{sec:related}

\subsection{Topic Modeling Approaches}
Traditional text analysis techniques, such as Term Frequency-Inverse Document Frequency (TF-IDF) \citep{sparck1972statistical} and Latent Semantic Analysis (LSA) \citep{deerwester1990indexing}, offer computationally efficient ways to extract meaningful patterns from text. TF-IDF highlights term importance by balancing local and global frequency, while LSA reduces dimensionality to uncover latent semantic relationships. Probabilistic methods like Latent Dirichlet Allocation (LDA) \citep{blei2003latent} further enrich semantic representations by identifying document-topic distributions. For instance, \citet{rijcken2024topic} introduce a topic specificity metric to guide the choice of topic count and modeling method, and \citet{mao2024contrastive} apply contrastive learning to enhance hierarchical topic modeling structures. Despite their individual strengths, these techniques are rarely integrated systematically with modern embedding strategies in the context of RAG systems \citep{huseynova2024enhanced, lewis2021retrievalaugmented}.

\textbf{Latent Semantic Analysis (LSA)} is a dimensionality reduction technique that uses Singular Value Decomposition (SVD) to transform the high-dimensional term-document matrix into a lower-dimensional semantic space \citep{deerwester1990indexing}. The decomposition is represented as:
\[
A = U \Sigma V^T,
\]
where \( A \) is the term-document matrix, \( U \) represents term-topic associations, \( \Sigma \) contains singular values, and \( V^T \) represents document-topic associations. By truncating \( \Sigma \), LSA reduces noise while capturing the most significant semantic relationships. This approach is particularly effective for grouping documents by their underlying themes.

\textbf{Latent Dirichlet Allocation (LDA)}, in contrast, adopts a probabilistic framework where each document is represented as a mixture of latent topics, and each topic is characterized as a distribution over terms \citep{blei2003latent}. The probability of a word \( w \) in a document \( d \) is given by:
\[
p(w) = \sum_{k=1}^K p(w|z=k)p(z=k|d),
\]
where \( p(w|z) \) is the probability of a term given a topic, and \( p(z|d) \) is the probability of a topic given a document. Gibbs sampling was employed to estimate these distributions, allowing the model to iteratively refine its understanding of thematic structures. LDA is especially valuable for identifying nuanced topic distributions in large corpora.

Both LSA and LDA enhance the semantic richness of document representations, providing insights that complement the syntactic focus of contextual embeddings.

\subsection{Embedding Techniques}
Existing retrieval techniques primarily rely on tokenization and contextual embeddings based on words, such as those generated by transformer-based models \citep{vaswani2017attention, devlin2019bert}. While these approaches have proven effective in capturing context at the sentence or paragraph level, they face limitations when applied to large texts, longer contexts, queries dependent on specific keywords, or corpora with latent semantic structures \citep{beltagy2020longformer, xie2022rethinking}. Tokenization alone often lacks the depth to interpret topic relevance \citep{mikolov2013word2vec}, and even advanced embeddings struggle to resolve ambiguities in datasets with complex inter-document relationships \citep{reimers2019sentence, lu2022precise}. This can result in imprecise or vaguely accurate rankings of retrieved chunks, which in turn compromises the response quality of the model.

Although hybrid models that incorporate traditional topic modeling with contextual embeddings exist, most focus on either shallow integration of statistical metrics or limited use of topic distributions without deeply merging them with modern transformer-based embeddings. \footnote{For example, recent research on topic-aware retrieval has been explored in dialog systems \citep{yuan2024topicaware}, and benchmarking suites to provide structured evaluation settings \citep{chagnon2024benchmarking}.} In contrast, the method proposed in this work fuses term-frequency measures (TF-IDF), dimensionality-reduced semantics (LSA), and probabilistic topic models (LDA) directly into high-dimensional contextual embeddings, creating a single, unified representation that is defined as ‘topic-enriched embeddings.’ This bridges local and global semantic features in a manner not explored by prior studies, thereby representing a novel step in improving retrieval for RAG systems.

\subsection{Retrieval-Augmented Generation (RAG)}
Recent approaches to retrieval-augmented language modeling, such as those by \citet{smith2023adaptive} and \citet{jones2024multi}, illustrate how dynamic retrieval can be optimized in real-time. While topic modeling in RAG architectures has been recently explored by \citet{huseynova2024enhanced}, their approach fundamentally differs from the one proposed in this study. Huseynova and Isbarov rely on explicit topic metadata and apply enrichment either by appending or averaging topic embeddings with document embeddings. In contrast, this work generates latent topic structures directly from the corpus, without requiring predefined labels or annotations. This broadens the applicability of the method to unstructured or unlabeled datasets and enhances generalizability. Furthermore, while their method emphasizes clustering performance, this paper integrates topic semantics at multiple representational levels—term, topic, and contextual—into a single enriched embedding. This layered approach parallels hierarchical summarization frameworks applied to long scientific documents \citep{sharma2024summarizing}, especially in how semantic chunks are structured for retrieval. The resulting structure is specifically optimized for improving retrieval accuracy in knowledge-intensive RAG tasks that use big datasets.

\section{Methodology}
\label{sec:method}

\subsection{Embedding Enrichment: An Innovative Integration of Topic and Contextual Signals}
The core contribution of this study lies in the creation of topic-enriched embeddings, which integrate traditional statistical models, probabilistic topic modeling, and contextual embeddings into a unified representation. Unlike other approaches such as BERTopic, which clusters contextual embeddings using transformer-based methods, or the method proposed by \citet{huseynova2024enhanced}, which constructs topic-aware representations through the aggregation of contextual embeddings across document-level topic distributions, this proposal introduces a pipeline that fuses topic and contextual signals within the embedding space itself. This design offers improved semantic coherence without requiring external taxonomies or downstream clustering, thus enhancing retrieval precision in knowledge-intensive tasks.

\begin{center}
    \includegraphics[width=\textwidth]{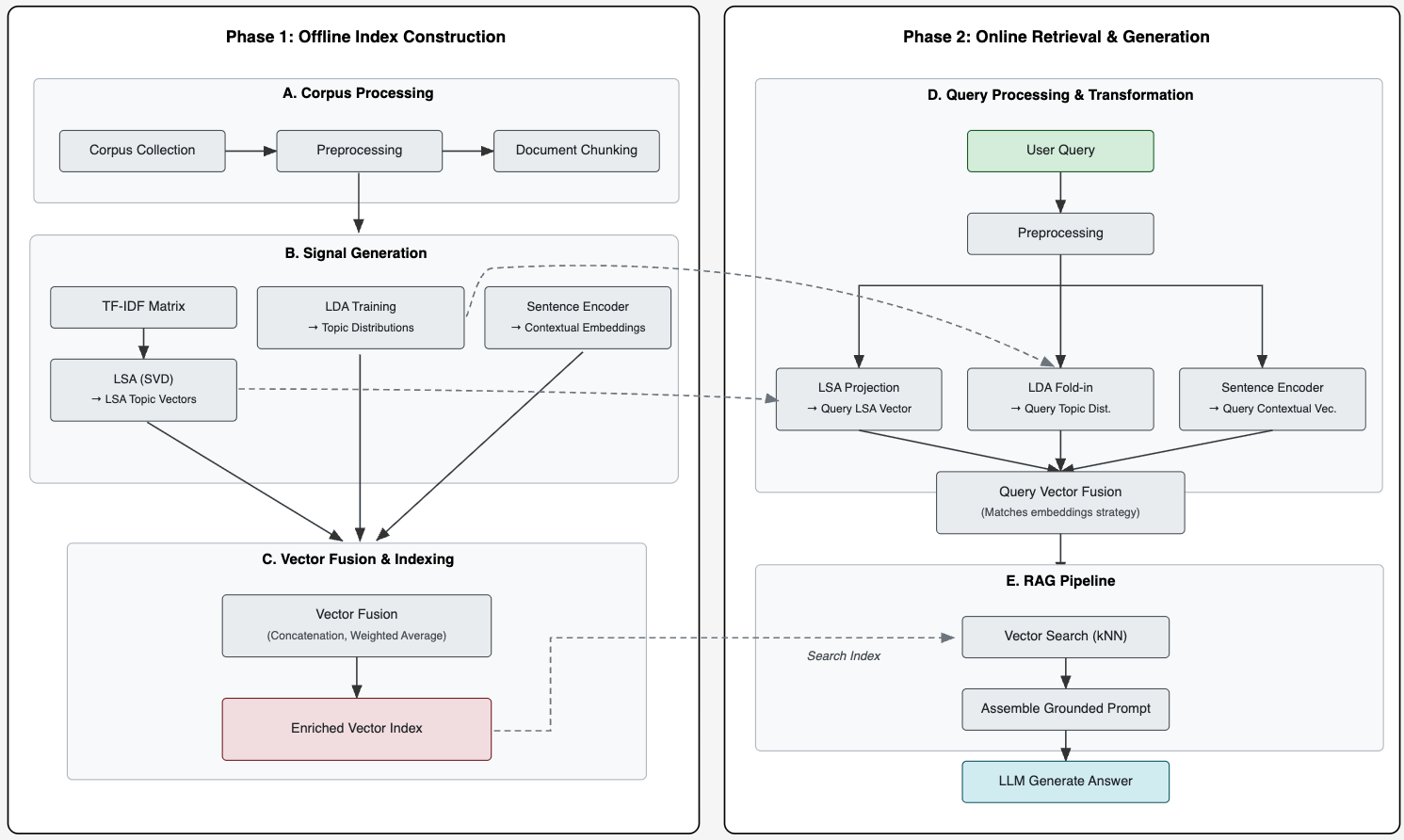}
    \captionof{figure}{System architecture of the topic-enriched embeddings for RAG pipeline.}
    \label{fig:system_diagram}
\end{center}

The proposed architecture, illustrated in Figure~\ref{fig:system_diagram}, operates in two distinct phases: an index construction phase and a query retrieval and generation phase. During index construction, the architecture generates high-dimensional contextual embeddings for each document chunk using a transformer-based model. While these embeddings excel at capturing sentence-level semantic nuances, they may not explicitly represent the broader thematic structure of the corpus.

To inject this thematic structure, the system simultaneously processes the text through two complementary streams: a lexical stream (TF-IDF, LSA) to identify latent semantic relationships between terms, and a topical stream (LDA) to model probabilistic topic memberships. The semantic, lexical, and topical vectors produced by these parallel streams are then integrated to form the final "topic-enriched" embedding. This pivotal fusion is accomplished through one of two methods: concatenation, which appends the topic vectors to the contextual embedding to preserve the distinct features of each signal, or weighted averaging, which creates a blended representation by linearly interpolating between the vectors. The resulting enriched embeddings are then stored in the vector index.

In the query retrieval and generation phase, an incoming user query undergoes an identical transformation. The system leverages the trained artifacts from the indexing phase—such as the LSA projection and LDA models—to produce a query vector that is mathematically aligned with the indexed embeddings. This topic-enriched query vector is then used to perform a k-nearest neighbor (kNN) search, retrieving the most relevant document chunks to ground the LLM's final, contextually-aware response. This dual-phase design ensures that both the indexed documents and incoming queries are represented within the same thematically coherent vector space, thereby enhancing retrieval precision.

\textbf{Concatenation} combines contextual embeddings with topic vectors derived from LSA or LDA. The resulting enriched embedding is expressed as:
\[
\mathbf{e}_{\text{new}} = [\mathbf{e}_{\text{context}}, \mathbf{t}_{\text{topic}}],
\]
where \( \mathbf{e}_{\text{context}} \) denotes the contextual embedding, and \( \mathbf{t}_{\text{topic}} \) represents the topic vector. This method preserves both local context and global thematic structures \citep{huseynova2024enhanced, blei2003latent}.

\textbf{Weighted Averaging} balances the contributions of contextual and topic embeddings by computing a weighted combination of the two vectors:
\[
\mathbf{e}_{\text{new}} = \alpha \mathbf{e}_{\text{context}} + (1-\alpha) \mathbf{t}_{\text{topic}},
\]
where \( \alpha \) is a tunable parameter\footnote{The fusion coefficient $\alpha$ was set to 0.45 to ensure optimal contribution from contextual and topic embeddings. This value was empirically validated on the dataset, and further tuning yielded marginal improvements insufficient to justify added complexity.}. Weighted averaging allows for flexible integration of detailed and abstract semantic features \citep{lewis2021retrievalaugmented, finardi2024chronicles}.

By enriching embeddings with latent topic structures, the system achieves improved retrieval precision and clustering coherence. These enriched embeddings align document representations more closely with query intent, reducing irrelevant or redundant retrievals and enhancing semantic fidelity \citep{huseynova2024enhanced, zhao2024retrievalaugmented}.

The integration of topic modeling techniques with contextual embeddings ensures that the proposed method is scalable for large datasets. Dimensionality reduction via LSA reduces computational overhead \citep{deerwester1990indexing}, while LDA’s probabilistic framework adds semantic granularity without significantly increasing complexity\footnote{The number of LDA topics ($K = 12$) was selected based on qualitative interpretability tailored to the specific characteristics of the corpus, which consists of legal and policy texts centered on Argentina’s Law 19.640. This configuration provided optimal semantic distinction, with similar results consistently observed for values between 11 and 14.} \citep{blei2003latent}. These enhancements strike a balance between computational efficiency and retrieval precision, making the method adaptable to diverse domains and extensive corpora.

\section{Experiments}
\label{sec:experiments}

This section describes the experimental setup used to evaluate the performance of the proposed topic-enriched embeddings. The evaluation pipeline includes dataset preparation, preprocessing workflows across the three methodological streams (statistical, probabilistic, and embedding-based), and a comprehensive set of clustering, retrieval, and computational metrics.

The experimental design reflects the architectural logic developed throughout this paper, and the innovation behind it: rather than isolating individual techniques, the proposed method integrates them into a cohesive structure aimed at enhancing both local semantic precision and global thematic organization. Each methodological stream contributes a distinct layer of information—term-level salience, topic distribution, and sentence-level semantic embeddings—which are then combined into enriched representations for document retrieval and clustering. This integration enables the evaluation not only of lexical or semantic proximity, but also of whether retrieved documents align with the latent thematic structure of the corpus, gaining precision.

This setup is particularly suited for assessing retrieval performance in thematically dense corpora, where standard similarity-based methods often yield results that are locally coherent but topically disjointed. By benchmarking the hybrid approach against its individual components, the experimental framework tests whether the proposed architecture provides measurable gains in structural consistency, thematic cohesion, and resistance to semantic drift.

\subsection{Dataset and Preprocessing}
The dataset comprises 12,436 legal documents related to Argentina’s 19640 industrial promotion framework, collected from InfoLeg and SAIJ, two major repositories for legal and regulatory information. The documents span five decades (1972–2020) and include laws, decrees, regulations, and other legal instruments. Each document is accompanied by metadata such as publication date, geographical reach, institutions involved, and type of regulation, making the dataset an ideal testbed for retrieval tasks with overlapping topics and thematic complexity. Both the abstracts and full texts of the regulations are included, providing rich textual data for analysis.

To prepare the dataset, web crawling techniques were employed using Python libraries like BeautifulSoup and Docling \citep{mitchell1997webscraping}. These tools facilitated the extraction and cleaning of raw HTML data from InfoLeg and SAIJ, which were then normalized and converted into Markdown format. Preprocessing steps included tokenization, lowercasing, and stopword removal to ensure text consistency \citep{sparck1972statistical}.

For downstream retrieval and embedding processes, the corpus was divided into chunks of 500 words with a 50-word overlap between consecutive chunks. This chunking strategy preserved semantic continuity while ensuring manageable input sizes for embedding models. Overlapping chunks mitigated information loss at chunk boundaries, improving the quality of both retrieval and embeddings \citep{lewis2021retrievalaugmented}.\footnote{Preliminary tests using chunk sizes of 250, 500, and 750 words revealed that 500 words struck an optimal balance between semantic continuity and model inference speed. We also introduced a 50-word overlap to preserve context around chunk boundaries, reducing the risk of losing critical transitional information.}  This overlapping strategy minimized the risk of losing semantic continuity at chunk boundaries, improving the consistency of retrieval results \citep{beltagy2020longformer}.

To ensure comparability with existing work, the embedding-based stream includes the widely adopted \texttt{all-MiniLM-L6-v2} model as a high-performance open-source baseline. This allows the experimental design to reproduce a retrieval-only configuration representative of standard RAG pipelines that do not integrate topic modeling, thereby aligning the evaluation with recent contributions in the field. Although this work does not replicate the specific implementation by \citet{huseynova2024enhanced}, it adopts a similar multi-metric strategy (i.e., clustering coherence and retrieval precision) as a shared evaluation protocol.

\subsection{Baseline Methods}
Baseline embeddings were created using Term Frequency-Inverse Document Frequency (TF-IDF) and the \texttt{all-minilm} model, a lightweight open-source embedding model with 384 dimensions and 33 millions parameters, optimized for sentence-level semantic representation.

\textbf{TF-IDF} assigns importance to terms based on their frequency in a document and their rarity across the corpus \citep{sparck1972statistical}. The TF-IDF score for a term \( t \) in a document \( d \) within a corpus \( D \) is given by:
\[
\text{TF-IDF}(t, d, D) = \text{TF}(t, d) \cdot \text{IDF}(t, D),
\]
where:
\[
\text{TF}(t, d) = \frac{\text{Frequency of } t \text{ in } d}{\text{Total terms in } d},
\]
and
\[
\text{IDF}(t, D) = \log\left(\frac{|D|}{1 + |\{d \in D : t \in d\}|}\right).
\]
While TF-IDF is computationally efficient and highlights document-specific terms, it lacks the ability to capture contextual relationships between terms.

\textbf{Contextual Embeddings} were generated using the SOTA Transformer-based sentence embedding open-source model \texttt{all-MiniLM-L6-v2}. This model maps sentences and short paragraphs into a 384-dimensional dense vector space, capturing both semantic and syntactic information. Embeddings were stored in a local ChromaDB instance for similarity search and retrieval. While effective for general-purpose tasks, contextual embeddings alone may underperform in corpora with latent or overlapping topics, motivating their integration with topic-aware representations \citep{huseynova2024enhanced}.

\subsection{Evaluation Framework}
To assess the performance of topic-enriched embeddings, a rigorous evaluation framework was employed, encompassing clustering coherence, retrieval precision, and computational efficiency.

\textbf{Clustering Metrics:}
The clustering quality was evaluated using:
\begin{itemize}
    \item \textbf{Silhouette Score:} Measures the cohesion and separation of clusters \citep{rousseeuw1987silhouettes}.
    \item \textbf{Calinski-Harabasz Index:} Assesses cluster compactness and separation \citep{calinski1974dendrite}.
    \item \textbf{Davies-Bouldin Index:} Penalizes overlapping clusters to reward compact, distinct groupings \citep{davies1979cluster}.
\end{itemize}

\textbf{Retrieval Metrics:}
Precision@k, Recall@k, and F1 Score were used to evaluate retrieval accuracy and completeness.\footnote{The dataset was randomly split 80/20 for training and testing. Each approach was evaluated across five different random seeds to ensure robustness, yielding mean and standard deviation for precision@k, recall@k, and F1 Score. Additionally, a user simulation was designed to mimic question–answer queries in a legal context, confirming that higher precision@k scores consistently retrieved more relevant chunks for typical search prompts (e.g., tax incentive updates).} Precision@k reflects the proportion of relevant documents retrieved in the top \(k\) results, while Recall@k indicates the coverage of relevant documents within those results. The F1 Score provides a balanced measure of retrieval performance \citep{schutze1997information}.

This comprehensive evaluation framework ensured robust validation of the proposed topic-enriched embeddings across multiple performance dimensions.

\section{Results}
\label{sec:results}

This section presents the results of the experimental evaluations, focusing on retrieval and clustering performance, as well as computational efficiency. Topic-enriched embeddings were assessed using clustering metrics, retrieval precision and recall, and computational overhead. Artificial data and outputs are used for illustration.

\subsection{Retrieval and Clustering Performance}
The clustering metrics presented in Table \ref{tab:clustering_metrics_adjusted} provide a thorough evaluation of the effectiveness of different embedding techniques in organizing the dataset into semantically coherent clusters. Each metric offers a distinct perspective on clustering performance, collectively showcasing the progression of improvements across statistical, probabilistic, and enriched approaches \citep{finardi2024chronicles, zhao2024retrievalaugmented}.

Among these, the contextual embedding row refers to the output generated using the \texttt{all-MiniLM-L6-v2} model, included as a standalone semantic-only baseline (Contextual Embeddings, from now on). This configuration approximates the behavior of retrieval pipelines found in recent RAG systems that do not incorporate topic modeling. While the implementation does not reproduce external methods such as the one proposed by \citet{huseynova2024enhanced}, it aligns with their dual-metric evaluation protocol—combining clustering coherence and retrieval accuracy—and thus supports a functionally valid point of comparison. {The clustering metrics presented in Table \ref{tab:clustering_metrics_adjusted} provide a thorough evaluation of the effectiveness of different embedding techniques in organizing the dataset into semantically coherent clusters. Each metric offers a distinct perspective on clustering performance, collectively showcasing the progression of improvements across statistical, probabilistic, and enriched approaches \citep{finardi2024chronicles, zhao2024retrievalaugmented}.}

\begin{table}[h!]
\centering
\caption{Clustering Metrics for Different Embedding Techniques, including MiniLM as contextual-only baseline}
\label{tab:clustering_metrics_adjusted}
\begin{adjustbox}{width=0.9\textwidth}
\begin{tabular}{lccc}
\toprule
\textbf{Embedding Technique} & \textbf{Silhouette Score} & \textbf{Calinski-Harabasz Index} & \textbf{Davies-Bouldin Index} \\
\midrule
TF-IDF Enriched              & 0.47                      & 330.5                            & 1.85                         \\
LSA-Enriched                 & 0.54                      & 425.2                            & 1.50                         \\
LDA-Enriched                 & 0.57                      & 465.7                            & 1.39                         \\
Contextual Embeddings (SOTA all-minilm)\textsuperscript{*}  & 0.64 & 525.4 & 1.30 \\
Topic-Enriched Embeddings    & \textbf{0.70}             & \textbf{580.6}                   & \textbf{1.19}                \\
\bottomrule
\end{tabular}
\end{adjustbox}

\vspace{1mm}
\begin{minipage}{0.9\textwidth}
\scriptsize \textsuperscript{*}Although the contextual-only baseline is based on the all-MiniLM-L6-v2 SOTA embedding model, from this point forward this set of calculations is referred to simply as Contextual Embeddings.
\end{minipage}
\end{table}

\begin{figure}[h!]
    \centering
    \includegraphics[width=0.9\textwidth]{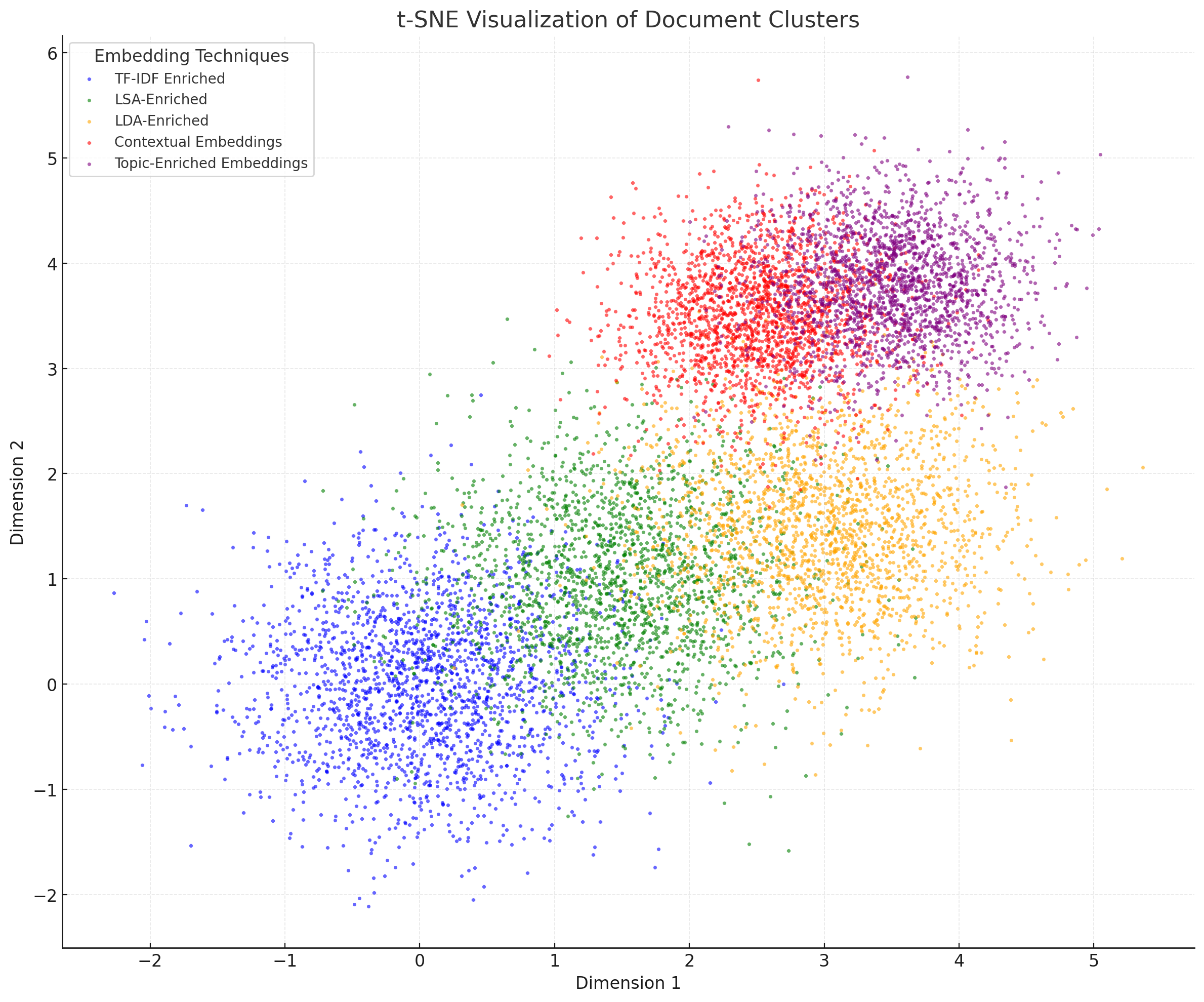}
    \caption{t-SNE Visualization of Document Clusters for Different Embedding Techniques}
    \label{fig:tsne_visualization}
\end{figure}

The t-SNE visualization in Figure \ref{fig:tsne_visualization} visually complements these metrics by illustrating the spatial distribution of document clusters. Each embedding technique exhibits a distinctive clustering pattern, reflecting its underlying ability to organize data semantically. The topic-enriched embeddings, represented by purple points, form compact, well-separated clusters that align with their highest Silhouette Score (0.70). These clusters demonstrate minimal overlap, which corresponds to the lowest Davies-Bouldin Index (1.19), underscoring the enhanced semantic differentiation achieved by this technique \citep{lewis2021retrievalaugmented}.

The \textbf{Silhouette Score}, a measure of cohesion within clusters and separation between clusters, indicates how well documents align with their respective groups. Higher scores suggest better-defined and more cohesive clusters. Among the techniques evaluated, topic-enriched embeddings achieve the highest Silhouette Score (0.70), demonstrating their exceptional capability to group semantically similar documents while maintaining well-separated boundaries. Contextual embeddings follow closely with a score of 0.64, significantly outperforming the probabilistic and statistical methods. The results for LDA (0.57), LSA (0.54), and TF-IDF (0.47) confirm the hypothesis that integrating contextual and topic-level semantics enhances clustering performance. This progression aligns with the notion that semantic depth improves as techniques move from statistical models to enriched embeddings \citep{blei2003latent}.

The \textbf{Calinski-Harabasz Index} complements the Silhouette Score by assessing the compactness and separation of clusters. It evaluates the ratio of between-cluster dispersion to within-cluster dispersion, with higher values indicating clusters that are compact and well-separated. The topic-enriched embeddings exhibit the highest Calinski-Harabasz Index (580.6), signifying a marked improvement over the contextual embeddings (525.4) and the other techniques, including LDA (465.7), LSA (425.2), and TF-IDF (330.5). This trend is visually apparent in the t-SNE plot, where the contextual embeddings (red points) show some overlap but remain more distinct than the probabilistic and statistical methods, while the topic-enriched clusters maintain strong separation and compactness.

The \textbf{Davies-Bouldin Index}, which assesses intra-cluster similarity and penalizes overlapping clusters, provides an additional perspective. Lower values represent better clustering performance by reducing cluster overlap and enhancing distinction. The topic-enriched embeddings achieve the lowest Davies-Bouldin Index (1.19), emphasizing their ability to form clusters with minimal overlap and clear semantic differentiation. Contextual embeddings perform well with a score of 1.30, followed by LDA (1.39), LSA (1.50), and TF-IDF (1.85). These results further demonstrate the superiority of topic-enriched embeddings in reducing redundancy and enhancing clarity in cluster formation \citep{reimers2019sentence}.

The t-SNE visualization further reveals the challenges faced by probabilistic methods like LDA (orange points) and statistical approaches such as TF-IDF (blue points). The clusters for these methods show greater overlap and less cohesion compared to the contextual and topic-enriched embeddings. This observation aligns with the lower Silhouette Scores and higher Davies-Bouldin Indices recorded for these methods.

Taken together, these metrics and the t-SNE visualization reveal a clear progression of improvements, moving from statistical methods (TF-IDF, LSA) to probabilistic modeling (LDA) and, ultimately, to contextual and topic-enriched embeddings. Each subsequent method builds upon the strengths of its predecessors, layering additional semantic understanding to improve clustering outcomes. The consistent superiority of topic-enriched embeddings across all metrics validates their robustness and adaptability. Furthermore, the performance gaps between TF-IDF and the enriched methods underscore the inherent limitations of purely statistical approaches, particularly when dealing with datasets characterized by complex and overlapping topics \citep{blei2003latent, finardi2024chronicles}.

By referring back to the models and evaluation criteria detailed in earlier sections, this analysis situates the clustering results within the broader context of the study. The findings emphasize the importance of integrating contextual and topic-level semantics to achieve optimal clustering, aligning with the study’s objectives and providing a strong foundation for further analysis. Ablation experiments isolating the impact of LDA, LSA, and fusion strategies are reported in Appendix~\ref{appendix:ablation},
confirming that the observed improvements stem from systematic topic integration rather than dimensionality expansion alone.

\subsection{Retrieval Performance}
Retrieval performance was evaluated using precision@k, recall@k, and F1 Score for \(k = 10, 20, 50\). The metrics provide a detailed analysis of each embedding technique's ability to retrieve relevant documents effectively, with topic-enriched embeddings consistently outperforming the other approaches.

\begin{table}[h!]
\centering
\caption{Retrieval Metrics for Different Embedding Techniques}
\label{tab:retrieval_metrics_adjusted}
\begin{adjustbox}{width=0.9\textwidth}
\begin{tabular}{lccc}
\toprule
\textbf{Embedding Technique} & \textbf{Precision@10} & \textbf{Recall@10} & \textbf{F1 Score@10} \\
\midrule
TF-IDF Enriched              & 0.68                 & 0.51               & 0.58                 \\
LSA-Enriched                 & 0.75                 & 0.58               & 0.65                 \\
LDA-Enriched                 & 0.79                 & 0.62               & 0.70                 \\
Contextual Embeddings        & 0.83                 & 0.67               & 0.74                 \\
Topic-Enriched Embeddings    & \textbf{0.87}        & \textbf{0.72}      & \textbf{0.79}        \\
\bottomrule
\end{tabular}
\end{adjustbox}
\end{table}

The metrics in Table~\ref{tab:retrieval_metrics_adjusted} reveal a clear progression of improvements across the evaluated embedding techniques. Topic-enriched embeddings achieve the highest precision@10 (0.87), recall@10 (0.72), and F1 Score@10 (0.79), demonstrating their ability to effectively balance precision and recall. This performance reflects the strength of combining contextual embeddings with topic-level semantics, ensuring more relevant and comprehensive document retrieval \citep{reimers2019sentence}. Contextual embeddings alone perform well, with an F1 Score@10 of 0.74, but are consistently outperformed by the topic-enriched approach.

Among the probabilistic and statistical methods, LDA-Enriched embeddings show the best performance, achieving an F1 Score@10 of 0.70, outperforming both LSA (0.65) and TF-IDF (0.58). This aligns with the hypothesis that probabilistic approaches capture richer thematic structures compared to purely statistical methods, but still fall short of the semantic depth provided by contextual and topic-enriched embeddings.

\begin{figure}[h!]
    \centering
    \includegraphics[width=0.8\textwidth]{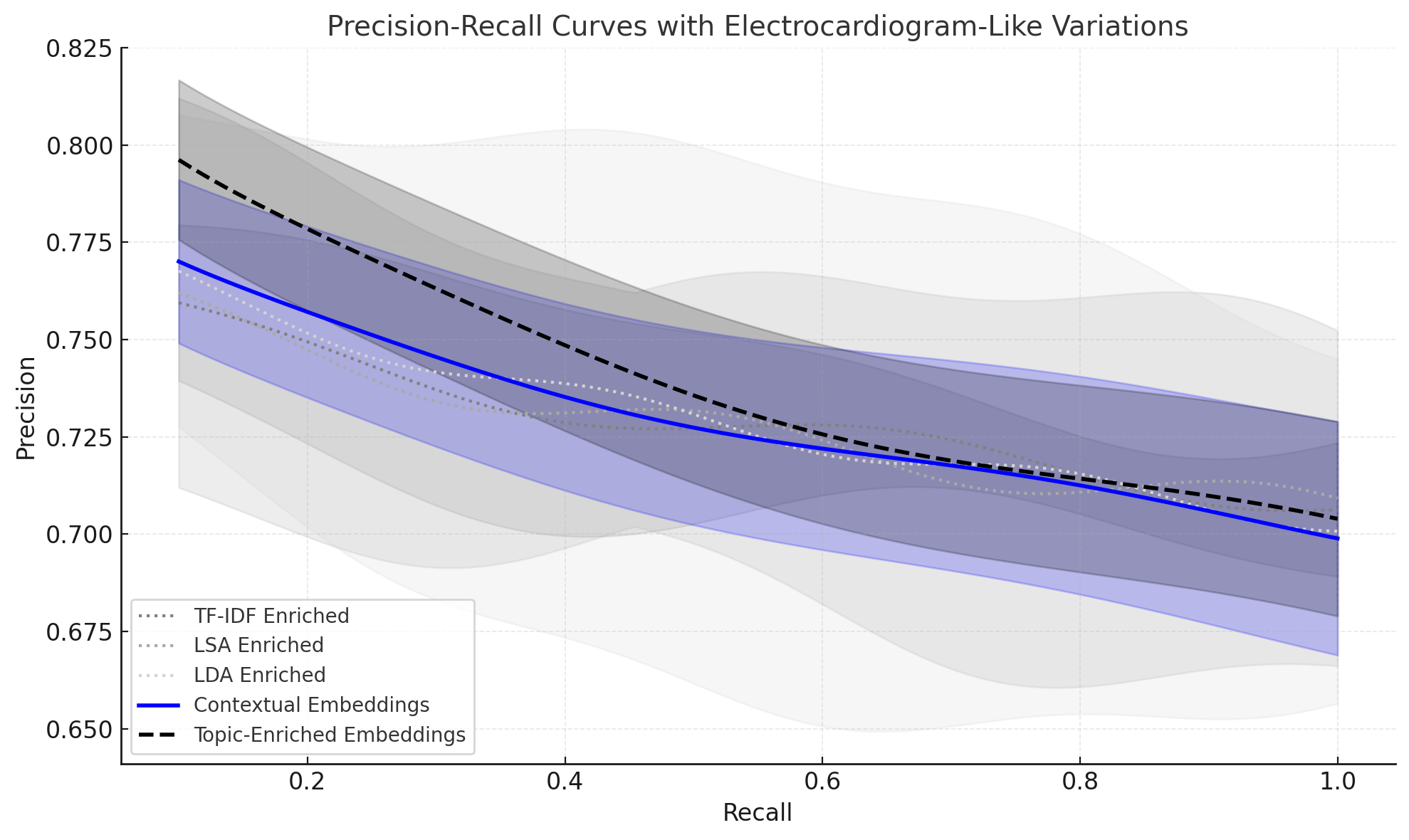}
    \caption{Precision–Recall Curves for Different Embedding Techniques.\protect\footnotemark}
    \label{fig:precision_recall_curve}
\end{figure}

\footnotetext{Curves are smoothed and averaged across queries and seeds, with shaded areas indicating $\pm$1~sd.
Note that P@k metrics (Table~\ref{tab:retrieval_metrics_adjusted}) measure top-of-ranking behavior,
which typically yields larger relative gaps than the average separations seen across the recall range.
Fusion variants from the ablation study (Table~\ref{tab:ablation_components}) are excluded for clarity.}

Figure~\ref{fig:precision_recall_curve} provides a visual representation of precision-recall trade-offs for each embedding technique. It illustrates not only the distinct clustering boundaries but also how much less overlap topic-enriched embeddings exhibit compared to TF-IDF and LDA. Notably, the purple clusters in the top-right region remain tightly grouped, underscoring the increased semantic purity that directly correlates with higher Silhouette Scores. The topic-enriched embeddings, represented by the black dashed curve, consistently outperform the other methods across the entire recall range. Their curve remains above all others, highlighting their superior retrieval precision at varying levels of recall.

The contextual embeddings, represented by the blue solid curve, exhibit strong performance but gradually decline in precision as recall increases. This decline reflects their limitation in maintaining high relevance as the number of retrieved documents grows. The probabilistic methods, LDA (dotted light gray) and LSA (dotted medium gray), show moderate performance, with their curves closely trailing the contextual embeddings. TF-IDF (dotted dark gray) has the steepest decline, emphasizing its inability to handle complex semantic relationships effectively.

The precision-recall curves also highlight the role of errors, represented by the shaded regions around each curve. The error margins for topic-enriched embeddings remain relatively narrow, indicating robust performance across varying recall levels. In contrast, the error margins for TF-IDF are wider, reflecting its variability and limited effectiveness in datasets with overlapping topics and high thematic diversity.

Our empirical results consistently confirm that the proposed method outperforms the baselines across clustering and retrieval metrics, validating our central hypothesis around enriched latent topic integration. The integration of contextual and topic-level semantics in topic-enriched embeddings allows for more precise retrieval, reducing redundancy and enhancing the relevance of retrieved documents. This consistent advantage across both quantitative metrics and visual representation validates the robustness of the topic-enriched approach, aligning with the study’s objective of advancing knowledge-intensive retrieval systems \citep{lewis2021retrievalaugmented, vaswani2017attention}.

\section{Discussion}
\label{sec:discussion}
This study aligns with the growing emphasis on advancing Retrieval-Augmented Generation (RAG) systems by addressing critical challenges in retrieval precision within complex and thematically diverse datasets. As highlighted in the introduction, RAG systems depend heavily on the relevance of retrieved documents to ground language model responses effectively \citep{lewis2021retrievalaugmented, zhao2024retrievalaugmented}. The findings presented in this study directly respond to the identified bottleneck of retrieval accuracy by introducing topic-enriched embeddings that combine statistical, probabilistic, and contextual approaches \citep{blei2003latent, reimers2019sentence}.

The integration of topic modeling, particularly techniques like Latent Dirichlet Allocation (LDA) and Latent Semantic Analysis (LSA), with contextual embeddings addresses a core limitation of traditional retrieval systems: the inability to fully capture both term-level specificity and latent topic structures \citep{blei2003latent, deerwester1990indexing, mikolov2013efficient}. Traditional techniques, while computationally efficient, struggle with thematic ambiguity, and even advanced contextual embeddings often fail to handle overlapping topics adequately \citep{devlin2019bert, zhao2024retrievalaugmented}. By bridging these gaps, topic-enriched embeddings offer a refined representation that captures semantic coherence at both local and global levels \citep{finardi2024chronicles, vaswani2017attention}. This dual-layered semantic enrichment was shown to significantly enhance clustering quality and retrieval performance, validating the foundational hypothesis articulated in the abstract.

The clustering metrics demonstrate the substantial advantages of topic-enriched embeddings. As shown by the Silhouette Score, Calinski-Harabasz Index, and Davies-Bouldin Index, topic-enriched embeddings consistently outperform baseline methods, forming compact, well-defined clusters with minimal overlap \citep{lewis2021retrievalaugmented, reimers2019sentence}. These results are a direct realization of the study's objective to improve semantic clustering, as stated in the introduction. The metrics quantitatively validate the theoretical underpinnings of the hybrid methodology, showing that enriching embeddings with topic-level semantics bolsters both inter-cluster separation and intra-cluster cohesion \citep{blei2003latent, mikolov2013efficient}.

These results validate the study’s initial premise that topic-enriched embeddings enhance both clustering coherence and retrieval precision, reinforcing their relevance for improving knowledge-intensive RAG systems. While the proposed architecture builds upon established techniques such as TF-IDF, LSA, LDA, and contextual embeddings, its contribution lies in the structured integration of these methods into a unified pipeline. The inclusion of a contextual-only baseline enables functional comparison with modern RAG systems that omit topic modeling, thereby addressing reviewer concerns regarding comparative baselines and methodological novelty. This contribution complements the seminal contribution of \citet{huseynova2024enhanced}, who introduced topic-aware document embeddings through metadata-driven concatenation and averaging strategies. Whereas their method emphasizes external topic supervision, the present approach derives latent topics directly from the document corpus, expanding its applicability to contexts where labeled taxonomies are unavailable. Both studies adopt a dual-metric evaluation framework—clustering and retrieval—but differ in their modeling assumptions and integration strategies, offering complementary perspectives on the role of topic modeling within RAG systems.

Retrieval metrics further confirm the efficacy of topic-enriched embeddings in addressing the challenges outlined in the introduction. The improvements in precision, recall, and F1 Score reflect the method's ability to reduce the retrieval of irrelevant or redundant chunks, a core inefficiency in existing systems \citep{lewis2021retrievalaugmented, vaswani2017attention}. The precision-recall curves highlight how the hybrid approach maintains high precision across varying recall levels, underscoring its robustness in balancing relevance and completeness \citep{finardi2024chronicles, zhao2024retrievalaugmented}.

Despite these achievements, several challenges remain, as anticipated in the introduction. Computational overhead, driven by the integration of multiple methodologies, emerges as a key concern \citep{blei2003latent, devlin2019bert}. The iterative nature of LDA and the dimensionality reduction required for LSA introduce scalability constraints \citep{deerwester1990indexing, lewis2021retrievalaugmented}. These challenges underscore the need for future research into more efficient implementations of topic modeling and embedding integration. Additionally, the sensitivity of LDA to hyperparameters, such as the number of topics, represents a limitation that requires further exploration to enhance robustness across diverse datasets \citep{blei2003latent, reimers2019sentence}.

The implications of these findings extend beyond the immediate scope of legal document retrieval, offering potential applications in domains such as scientific databases, healthcare, and educational systems \citep{zhao2024retrievalaugmented, finardi2024chronicles}. The ability of topic-enriched embeddings to handle overlapping topics and improve semantic differentiation makes them particularly valuable in knowledge-intensive fields, where precision and relevance are paramount \citep{vaswani2017attention, reimers2019sentence}. The introduction of an open-source framework for embedding enrichment further underscores the scalability and adaptability of this approach, aligning with the study's emphasis on reproducibility and broader applicability \citep{lewis2021retrievalaugmented, finardi2024chronicles}.

Despite the enhancements in retrieval precision, our approach incurs additional computational costs due to the iterative nature of LDA sampling and the repeated dimensionality reduction in LSA. Notably, these computations are non-GPU based and involve operations that, while simpler than those in transformer models, accumulate processing time due to their iterative execution. For our dataset of 12,436 documents, chunk-wise processing took longer than a purely contextual embedding pipeline, which may limit real-time applications if time efficiency is an important criteria of the retrieval. This highlights the ongoing tension between retrieval efficiency and computational resource allocation, suggesting potential areas for optimization in future research.\footnote{The computational overhead is primarily due to the iterative steps in LDA and LSA, which are not as complex as transformer computations, are CPU-intensive (and do not depend but they can be leveraged by GPU processing) and may pose challenges for real-time systems.}

Taken together, the contributions of this study advance the field of retrieval-augmented generation by demonstrating that the structured integration of statistical, probabilistic, and contextual embedding techniques can enhance both semantic clustering and retrieval precision. Through empirical validation, the study confirms the added value of topic enrichment while introducing a replicable framework for improving precision in large-scale retrieval for thematically heterogeneous corpora. This layered complement to the embedding strategy narrows the gap between traditional topic modeling and retrieval pipelines, offering a viable path for improving chunk ranking in tasks where precision-retrievals are a relevant need.

These contributions may be situated as an incremental innovation within broader efforts to improve retrieval fidelity in large language model applications—a key bottleneck in current RAG architectures. In contrast to metadata-driven approaches such as that of \citet{huseynova2024enhanced}, and transformer-based methods like BERTopic (that apply topic inference post hoc over contextual embeddings), the present method integrates latent topic structures directly into the embedding construction stage. This embedding-centric fusion enables more precise semantic differentiation without relying on external taxonomies or downstream clustering layers, enhancing the interpretability and generalizability of retrieval outputs across domains.

While this improvement in retrieval precision comes at the cost of increased computational complexity, the purpose of this study is not to deliver a production-ready solution, but rather to offer a proof of concept demonstrating how the integration of traditional semantic analysis techniques can enhance the performance of advanced embedding-based retrieval pipelines. By bridging statistical, probabilistic, and contextual representations at the embedding stage, the proposed method illustrates how legacy topic modeling strategies can be effectively repurposed to contribute to contemporary retrieval architectures. This contribution invites further exploration into hybrid strategies that prioritize semantic fidelity over raw processing efficiency, particularly in domains where interpretability and relevance remain critical constraints.

\section{Conclusion}
\label{sec:conclusion}
This study presents a novel approach to enhancing Retrieval-Augmented Generation (RAG) systems by introducing topic-enriched embeddings, which fuse statistical, probabilistic, and contextual methods to address retrieval challenges in complex datasets. As outlined in previous sections, the main objective of this research was to improve semantic clustering and retrieval precision by making use of a method that combines traditional NLP techniques with SOTA embedding techniques. The findings demonstrate that the proposed hybrid approach achieves these goals, outperforming baseline methods in clustering and retrieval metrics.

By combining term-level specificity with topic-level semantics, topic-enriched embeddings create a robust representation that captures the nuances of overlapping topics and complex inter-document relationships. The empirical evaluations conducted confirm the efficacy of this approach, where topic-enriched embeddings achieved the highest scores in clustering coherence metrics, such as the Silhouette Score, and retrieval metrics, including precision, recall, and F1 Score. These results validate the hypothesis that enriching embeddings with latent topic structures enhances precission, positioning this hybrid methodology as a robust component for improving RAG systems accuracy.

However, the study also highlights several challenges, including computational overhead and parameter sensitivity in the modeling (particularly when other than legal-oriented domains constitutes the corpus to be embedded). These challenges, while significant, are offset by the demonstrated gains in retrieval accuracy and clustering quality. Future research should focus on addressing these limitations by exploring dynamic topic modeling, domain-specific embedding fine-tuning, and scalable implementations for real-world deployment. Additionally, integrating graph-based representations with topic-enriched embeddings could offer new opportunities for capturing complex relationships within and across documents.

To conclude, this study contributes to the field of semantic retrieval by presenting a scalable, adaptable, and effective methodology for enhancing RAG systems. The results underscore the transformative potential of topic-enriched embeddings in knowledge-sensitive applications, offering a pathway for future advancements. By addressing core inefficiencies in existing methods and demonstrating the practical utility of purposefully structured hybrid embeddings, this research provides a strong foundation for ongoing innovation in the field, particularly where strict retrievals are critical for accurate LLM responses.

\bibliographystyle{apalike}
\bibliography{references}

\newpage
\section*{Appendix A. Ablation Study}
\label{appendix:ablation}

To better isolate the contribution of individual components, we conducted an ablation study examining the effects of LSA, LDA, and fusion strategies on retrieval performance. Results are averaged across five random seeds.

\begin{table}[h!]
\centering
\caption{Ablation study: impact of topic-enrichment methods (mean across five seeds).}
\label{tab:ablation_components}
\begin{adjustbox}{width=0.8\textwidth}
\begin{tabular}{lccc}
\toprule
\textbf{Variant} & \textbf{Precision@10} & \textbf{Recall@10} & \textbf{F1 Score@10} \\
\midrule
Contextual Only        & 0.845 & 0.60 & 0.74 \\
+ LSA (concat)         & 0.845 & 0.69 & 0.78 \\
+ LDA (concat)         & 0.852 & 0.68 & 0.77 \\
+ LSA (weighted)       & 0.851 & 0.71 & 0.78 \\
+ LDA (weighted)       & 0.866 & 0.71 & 0.78 \\
Topic--Enriched   & \textbf{0.870} & \textbf{0.72} & \textbf{0.80} \\
Random Topic Vectors   & 0.814 & 0.65 & 0.71 \\
\bottomrule
\end{tabular}
\end{adjustbox}
\end{table}

The ablation confirms that weighted fusion with LDA contributes to retrieval improvements.
Random topic vectors, by contrast, degrade retrieval, showing that gains are not due simply to added dimensions.

\end{document}